\documentclass[10pt,conference]{IEEEtran}
\IEEEoverridecommandlockouts
\usepackage{cite}
\usepackage{amsmath,amssymb,amsfonts}
\usepackage{graphicx}
\usepackage{textcomp}
\usepackage{xcolor}
\def\BibTeX{{\rm B\kern-.05em{\sc i\kern-.025em b}\kern-.08em
    T\kern-.1667em\lower.7ex\hbox{E}\kern-.125emX}}

\usepackage[utf8]{inputenc}
\usepackage{color}
\usepackage[T1]{fontenc}
\usepackage[english]{babel}
\usepackage{graphicx}

\usepackage{url}

\usepackage{listings}

\usepackage{authblk}

\usepackage{caption}
\usepackage[official]{eurosym}

\usepackage{cleveref}
\crefname{section}{§}{§§}
\Crefname{section}{§}{§§}

\usepackage{setspace}

\begin{document}

\title{DNS-based dynamic context resolution for SCHC}%Resolving SCHC Context using DNS}%Using the DNS infrastructure for resolving Context in IoT\\Alternative title :  \\Using the DNS infrastructure to query device-specific information in IoT}% : Application to SCHC compression in LoRa Networks}

%\author{Antoine BERNARD, Sandoche BALAKRICHENAN, Michel MAROT}
\author[1,2]{Antoine Bernard}
\author[1]{Sandoche Balakrichenan}
\author[2]{Michel Marot}
\author[1]{Benoit Ampeau}

\affil[1]{AFNIC}
\affil[ ]{\textit{firstname.surname@afnic.fr}}
\affil[2]{Samovar, CNRS UMR 5157, Télécom SudParis, Institut Polytechnique de Paris}
\affil[ ]{\textit{firstname.surname@telecom-sudparis.eu}}

\maketitle

\begin{abstract}
LPWANs are networks characterised by the scarcity of their radio resources and their limited payload size. LoRaWAN offers an open, easy-to-deploy and efficient solution to operate a long-range network. To efficiently communicate using IPv6, the lpwan working group from the IETF developed a solution called Static Context Header Compression (SCHC). It uses context rules, which are linked to a given End Device, to compress the IPv6 and UDP header. Since there may be a huge variety of End Devices profile, it makes sense to store the rules remotely and use a system to retrieve the profiles dynamically. In this paper we propose a mechanism based on DNS to find the context rules associated with an End Device and, allowing it to be downloaded from an HTTP Server. We evaluate the corresponding delay added to the communications using experimental measurements from a real testbed.
\end{abstract}

\textbf{Keywords - Internet of Things; LPWAN; SCHC; DNS}

\section{Introduction}
\label{section:intro}

Internet of Things (IoT) includes a wide range of technologies relying on a variety of standards and protocols. Each of the IoT technologies is suited for different IoT use cases. One of the main limitations of IoT technologies, when compared to the Internet, is that most of IoT devices, as well as the network, have constrained capabilities.

Low-Power Wide-Area Networks (LPWANs)\cite{rfc8376} is one such IoT technology that aims to provide network connectivity to IoT devices distributed over a wide area. Their distinct characteristics such as their coverage capacity (ranging from ten to fifteen kilometers), long battery life (with a lifespan of more than ten years), low cost (the architecture reduce expensive infrastructure requirements, and the use of license-free or already owned licensed bands reduce network costs) satisfies the requirements of a considerable IoT market \cite{iot_market}. Major LPWAN technologies include Long Range Wide Area Network (LoRaWAN) \cite{lorawan}, Narrow Band IoT (NB-IoT) \cite{nb-iot}, LTE-MTC (Machine Type Communication, LTE-M) \cite{lte-m}, Sigfox \cite{sigfox}, Wi-SUN Alliance Field Area Network (FAN) \cite{wi-sun_fan}.

Most LPWAN technologies use a star topology network. Data sent from the End-Device (ED) is received by a Radio Gateway (RG) which relays the data to a Network Server (NS). The NS relays the data and additionnal information (signal quality, channel, remaining lifetime, LoRa version, device version) to the appropriate Application Server (AS). Usually, the RG, the NS and the AS are either connected via a direct IP link or connected to the Internet.

The communication bandwidth in the LPWAN, that is the radio connectivity between the ED and the RG, is highly constrained. Because of the scarcity of resources and duty cycle constraints, data compression has become a necessity.

Static Context Header Compression (SCHC) \cite{schc} is a framework that provides both compression and fragmentation functionalities. It is being standardised by the lpwan \cite{lpwanwg} working group at the IETF. The SCHC document is still in the draft format and is in the final process to become a Request For Comments (RFC). The SCHC draft and its related works \cite{lpwanwg} at the IETF is also being followed by other standards developing organizations such as IEEE 802.15, or the LoRa Alliance. It is considered an efficient solution to connecting the LPWANs using IPv6, thus enabling end-to-end IP connectivity. With the help of the SCHC framework, it is possible to compress an IPv6 header from its original size of sixty bytes down to two bytes, thus reducing bandwidth usage and increasing communication efficiency.

In this paper, we propose a remote context querying mechanism to allow application providers and developers to easily and quickly manage context data using DNS queries and use SCHC's functionalities on the server side. Even though the proposed mechanism has not been tried earlier to our knowledge, in \cref{section:SotA} we provide the motivation for this research as well as related work involving SCHC.
In \cref{section:Experiment}, we explain our methodology and our experimental testbed and finally in \cref{section:Results}, we analyze our results. 

\section{Motivation and Related works}
\label{section:SotA}

As proven by \cite{ugent}, SCHC provides "a unified protocol stack independent of the underlying LPWAN technology". This proves particularly useful when designing product with multiple connectivity, allowing developers to communicate on LPWANs using the IP protocol.
With the evolution of the IoT market, the only reliable and efficient solution to connect a device will be using IPv6, but IPv6 packets are too heavy to be sent on LPWANs without adaptation. SCHC provides such adaptation, but also saves a lot of development time from technology specific adaptations.
For example, \cite{dlms_whitepaper} provides specification for the transport of DLMS messages over LPWAN using SCHC, which saves time for their application developers and protocol designers while fulfilling security considerations.

\begin{figure}[ht!]
    \centering
    \includegraphics[width=\linewidth]{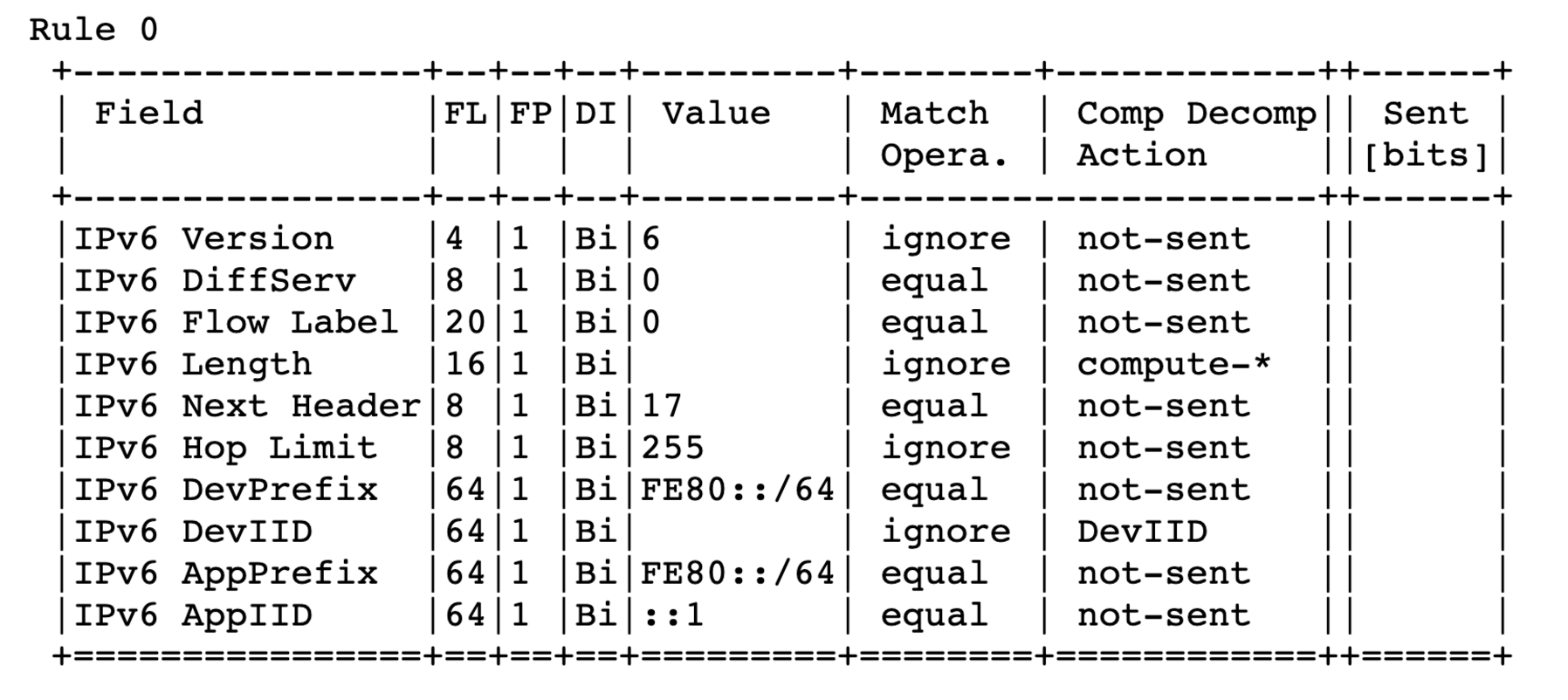}
    \caption{Context rule example (Source \cite{schc})\label{schc_rule}}
\end{figure}

In order to compress the data sent and received between the ED, the NS and the AS, SCHC uses a predefined group of rules called {\bf Context} which is deployed on the ED and on the NS or the AS. For our experiments, the Context is deployed at the ED and the AS. 
This context may be specific for each ED or common for a group of EDs. Fig. \ref{schc_rule} is an example of Context Rule as described in the SCHC draft \cite{schc}. 

For every Context, there could be single or multiple rules. When sending data from the ED to the RG, the SCHC Context rule enables compression by suppressing redundant, superficial, predictable and/or most used data inside an IPv6 header and replacing them with a Rule Identifier chosen in a given set of predefined rules. For instance, the ED's IP address may be added to the Context allowing to avoid transmitting 128 bits IP address data if all the packets sent by a sensor have the same IP address. When using SCHC, either the RG, the NS or the AS is supposed to realize SCHC operation (compression, decompression, fragmentation, reassembly) for all associated EDs.
If the Context rules are hard-coded in the ED, the NS or the AS, it will be challenging to modify later. That is the reason why we propose that the ED, the NS or the AS should retrieve the Context dynamically from a remote server. Thus, the owner of the rules could easily modify them. Only the Rule ID’s are stored either in the ED, the NS or the AS.

Managing 15 rules stored on a single device is possible, but the system would face trouble when up-scaling as each rule is uniquely bound to its device. With around 10.000 devices around a single radio gateway, and multiple radio gateway for a network gateway, we suddenly end up with potentially a lot of information to store. If we use, for example, Chirpstack Gateway OS \cite{chirpstackgatewayos} which is designed to work on a Raspberry Pi and embarks all the software necessary to operate a LoRa Network, it would be costly to store the rules for all the devices around the gateway

There are multiple options of storage for Context rules. It could be done in a private server, stored in the cloud or provided by a third party application. However we think that it could be wise to use a standardized open, distributed and easy to use mechanism to find the location of the server where the Context Rules are stored. . Thus, we zeroed in on DNS \cite{rfc1034} \cite{rfc1035} , which is the only optimized hierarchically distributed database which could enable to identify the location of the server where the Context Rules are stored in a feasible and efficient manner on the Internet.

To our knowledge, there is only another work \cite{schc_roaming}, which tries to retrieve the Context from an Administration Management Server (AMS) for roaming purposes in LoRaWAN. Our opinion is that the AMS scenario is operationally possible only when the operators have a contract between each other in order to retrieve the location of the AMS. But the advantage of LoRaWAN is that with an investment of less than 500 \euro , one can set up a private LoRaWAN connection. Thus, there could be thousands of private networks for three or four public networks (where large operators cover a whole Country for instance) in some geographical area. The only difference being that private networks use a default network identifier, open to everyone and allowed for private or research use. Though they are private, these network are still connected to the internet and may still be contacted from the outside. But to have contracts between thousands of private/public network operators for resolving the location of the AMS is not operationally feasible. That is the reason we zeroed in on using the DNS for resolving the location of the server that hosts the Context rules.

%we chose not to focus on transmitting the rules to the devices over the air and considered that the rules would already be on the devices as the current standard prosposes it.

%%%%%%
%
% Need more data on SCHC implementations
%
%%%%%%

\section{Proposed Mechanism and experiment}
\label{section:Experiment}

Our experiment aims to test if SCHC-enabled communication is viable in a real use-case regardless of its theoretical feasibility. Then, we want to test if SCHC remote rule-management is possible and propose a DNS-based mechanism to support SCHC remote rule management. We provide measurements to study the consequences of a remote context retrieval system on latency in LoRa networks (for server-side services and for devices). Our experimental results also aim to serve as references to enable further work on the subject.

\begin{figure}[ht!]
    \centering
    \includegraphics[width=\linewidth]{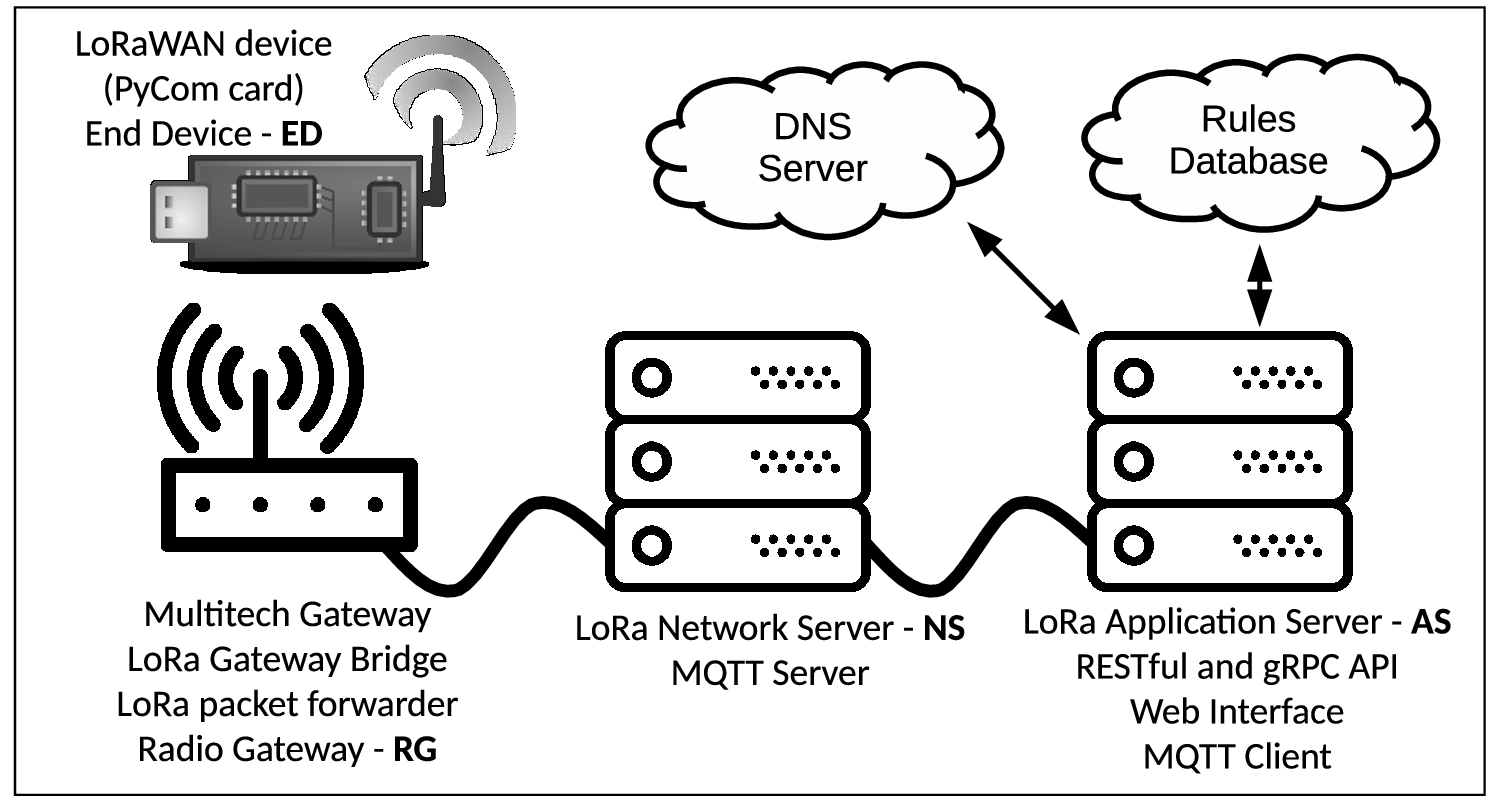}
    \caption{Measurement Platform's Architecture \label{lora_platform}}
\end{figure}

The proposed mechanism consists in delocalizing the SCHC rules on a remote HTTP server and using DNS to retrieve them. When the ED sends data, they are received by the RG, then transferred to the NS and retrieved by the AS. To decompress the data, the AS needs access to the rule. We use the DNS to retrieve a hash of the rule (since it is not possible to store the whole rule in the DNS for now) and possibly an address of the HTTP server on which the corresponding rule might be stored.
We can safely assume for our experiment that the rules can be uniquely linked to the tuple (DevEUI, RuleID). 

This tuple is constructed by extracting the DevEUI from the LoRa frame and the RuleID using the first bits from the LoRa payload compressed by SCHC. The AS is used to store the rules corresponding to the device in its perimeter in a rules cache for a set period of time. The rules in the AS are indexed in a hash table. When the AS receives data, it constructs the tuple (DevEUI, RuleID) as indicated above, then uses the DNS to retrieve the hash of the corresponding rule and search for this rule in its rules cache. If it is not found because it is a new tuple (DevEUI, RuleID), a new rule must be stored in the cache and the HTTP server where the rule is stored is interrogated to get it. Then, the rule is inserted in the cache. Note that even when a rule is present in the cache, the DNS is systematically queried because the freshness of the information must be checked to ensure that the rule has not been modified since its last cache insertion.
Finally, the data can be decompressed. Once the data are decompressed, a response may be sent back by the server depending on the needs of the application. Fig. \ref{lora_platform} presents the interactions between the AS, the DNS and the HTTP server in the case where a new rule is needed.

\subsection{Measurement scenarios}

Our study focuses on Application Server Response Time and End Device Uplink Round Trip Time through different scenarios. Scenario 1 and 2 serve as references to compare with other scenarios. They provide the minimum communication time with and without SCHC decompression. Scenario 3 aims to study the mechanism presented in \cref{section:Experiment}. Scenario 4 studies the case where most of the information is always present in the cache. These four scenarios are described more precisely below: \begin{itemize}
\begin{figure}[ht!]
    \centering
    \includegraphics[width=\linewidth]{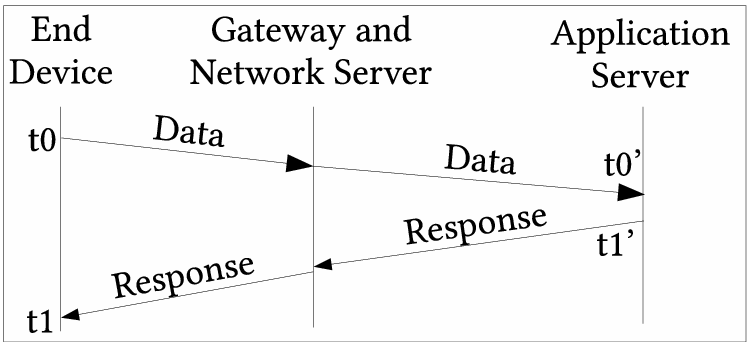}
    \caption{Message Exchange Diagram (Scenario 1)\label{ex_diag_no_decomp}}
\end{figure}
\item Scenario 1: The first measurement is designed to be used as an experimental reference for our platform. Data are sent without compression from the ED over LoRa and a response is sent back from the LoRa AS in order to measure the RTT $t1-t0$ (cf. Fig. \ref{ex_diag_no_decomp}). We also measure the Application Server Response Time $t1' - t0'$. No decompression operation is performed on the data. 
\begin{figure}[ht!]
    \centering
    \includegraphics[width=\linewidth]{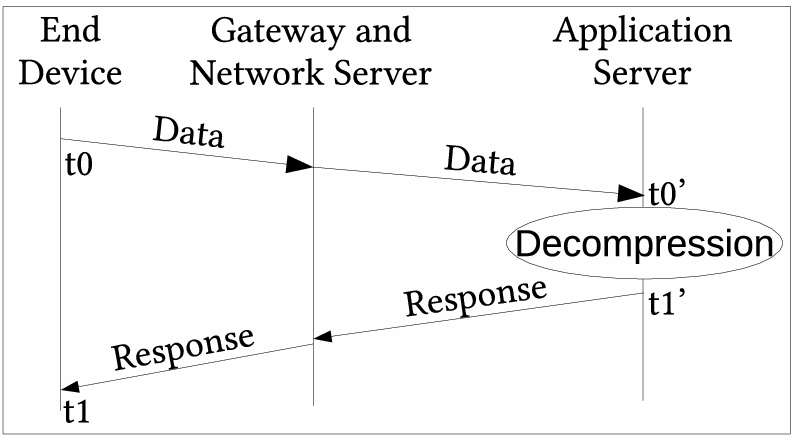}
    \caption{Message Exchange Diagram (Scenario 2)\label{ex_diag_local}}
\end{figure}
\item Scenario 2: The second measurement adds the SCHC mechanism for the communication over LoRaWAN. The ED sends the data compressed using the SCHC Context and the received data are decompressed using the same Context rule that is stored in a file locally on the AS. We measure $t1-t0$ (cf. Fig. \ref{ex_diag_local}) We also measure the Application Server Response Time $t1' - t0'$. The comparison with results from Scenario 1 allows us to estimate the decompression time.
\begin{figure}[ht!]
    \centering
    \includegraphics[width=\linewidth]{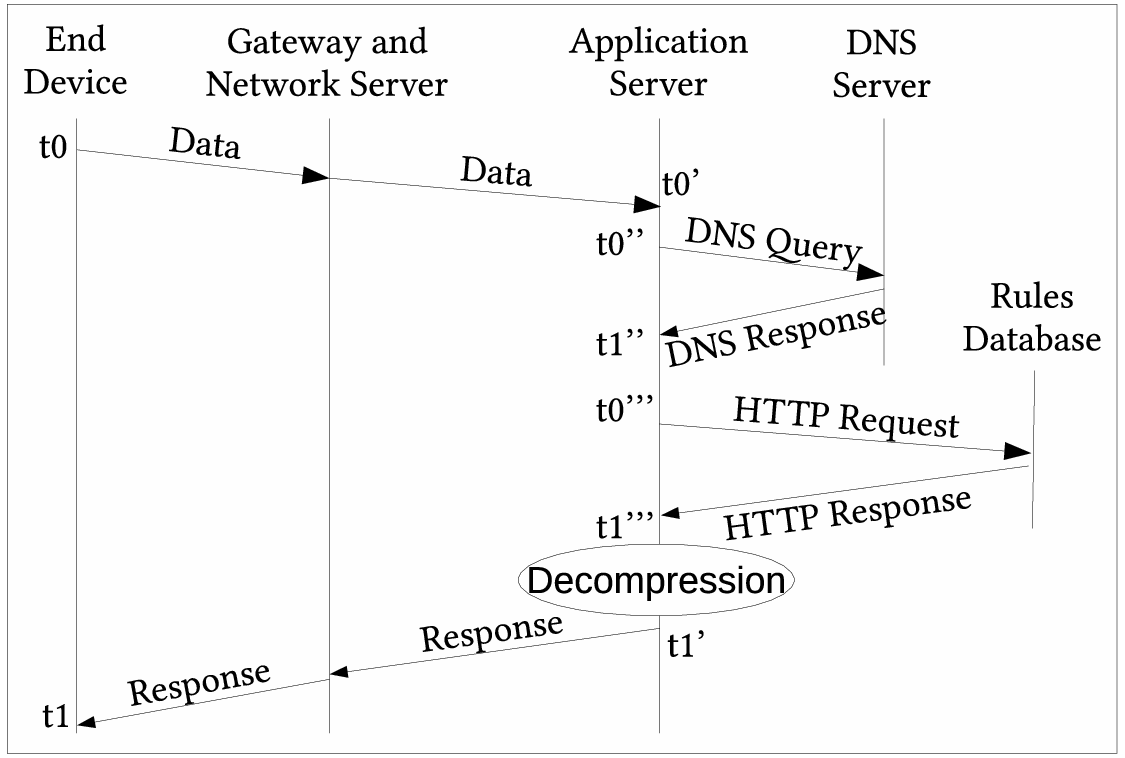}
    \caption{Message Exchange Diagram\label{ex_diag_all}}
\end{figure}
\item Scenario 3: The third measurement is the key scenario of our study. It aims to add the mechanism presented at the beginning of \cref{section:Experiment} and illustrated by Fig. \ref{lora_platform} to provide the AS with the SCHC Context that is stored in a remote server. In this measurement, instead of using a locally stored Context rule for decompression, the AS is asked to download the Context file from a remote HTTP server with a request such as "HTTP GET myschcrules.net/DevEUI/RuleID". We measure the total Round Trip Time (RTT) $t1 - t0$, the Application Server Response Time $t1' - t0'$, the RTT of the DNS Query $t1'' - t0''$ and the RTT of the HTTP Request $t1''' - t0'''$ (cf. Fig. \ref{ex_diag_all})
\begin{figure}[ht!]
    \centering
    \includegraphics[width=\linewidth]{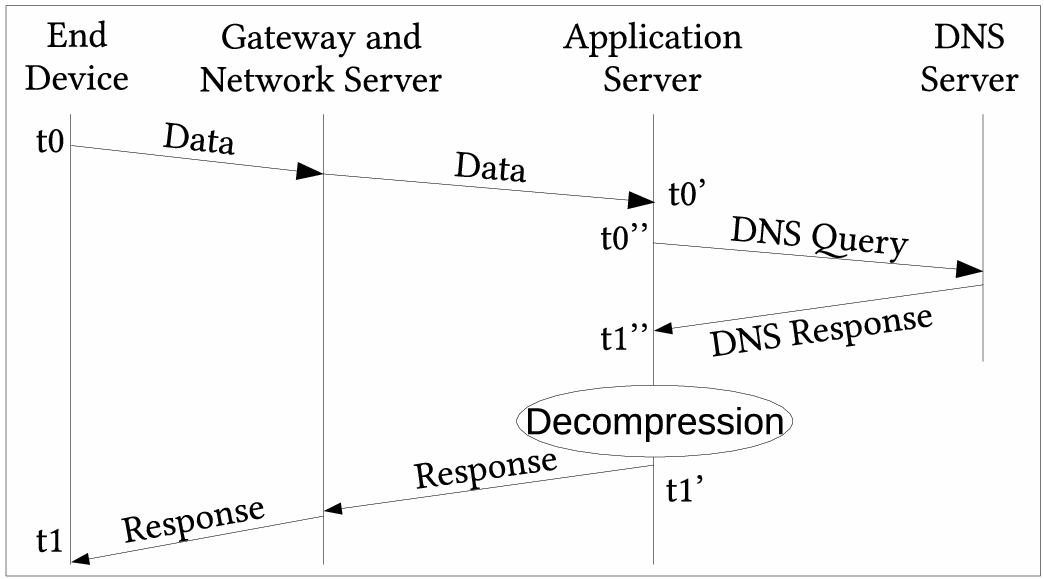}
    \caption{Message Exchange Diagram (Scenario 4)\label{ex_diag_dns}}
\end{figure}
\item Scenario 4: In most cases, EDs will be static (e.g. Water Meters) and well known by the AS so their context rules will be always present in the AS cache of rules. In this case, the DNS is still queried at least to check there has been no change to the rule. This scenario 4 corresponds to the case where there is no changes to the rule, the rule has not been updated and it is still present in the AS cache. Thus there is no need to query the HTTP server and download the rule again. We measure the total RTT $t1 - t0$, the Application Server Response Time $t1' - t0'$ and the RTT of the DNS Query $t1'' - t0''$ (cf. Fig. \ref{ex_diag_dns}).
\end{itemize}

To sum up, scenario 1 and 2 are used as experimental references, scenario 3 will be the case for initial communications, rules update and cache expiration whereas scenario 4 would correspond to day-to-day queries for Device Context Status.

\subsection{Experiment Testbed}

As mentioned earlier, LoRaWAN is a LPWAN technology. It uses an open standard and is easy to deploy so we use it for our experiments. ChirpStack\cite{loraserver} is an open-source solution to easily build a ready-to-use LoRaWAN. It provides the software components of our infrastructure for the RG, NS and the AS.

ChirpStack works with the RG to ensure that the data received from the devices can be relayed to the AS. For our experiment, we chose to connect directly to a MQTT broker and subscribe to the message queue associated to our devices, but MQTT can also be used to monitor the gateway or to contact all the devices that are linked to a specific LoRa Application using various topics. LoRa AS also offers a REST API, a gRPC API and a web UI to offer multiple ways to operate a LoRaWAN network.

We used PyCom FiPy development cards as LoRa-enabled devices and we made them send SCHC compressed data based on a Context over LoRa to a Multitech Conduit RG which forwards the data to the ChirpStack Network Server. Then we can retrieve the data using the ChirpStack Application Server or subscribe to the MQTT broker hosted on the Network Server to retrieve the data sent and decompress it based on the same Context that was used for compression.

The SCHC implementation we used to decompress data is OpenSCHC \cite{openschc}. This implementation is developed by the authors of the SCHC internet draft as a proof of concept. It serves as the base reference for other SCHC implementations.

FiPy cards are Class A compliant devices as defined by the LoRa Standard \cite{lorawan}, hence they respect a strict emission/reception schedule. Our experimentation is realised respecting the EU regulations on duty cycle, communicating in the EU 868 MHz frequency. All communications are done using Spreading Factor (SF) 7 considering that, for our experiment, it is the one we expect to include most constraints regarding latency. If our system works without hindering RTT for SF7, it has no reason to hinder the RTT for higher latency SF.

\section{Results and discussion}
\label{section:Results}

\begin{figure}[ht!]
    \centering
    \includegraphics[width=\linewidth]{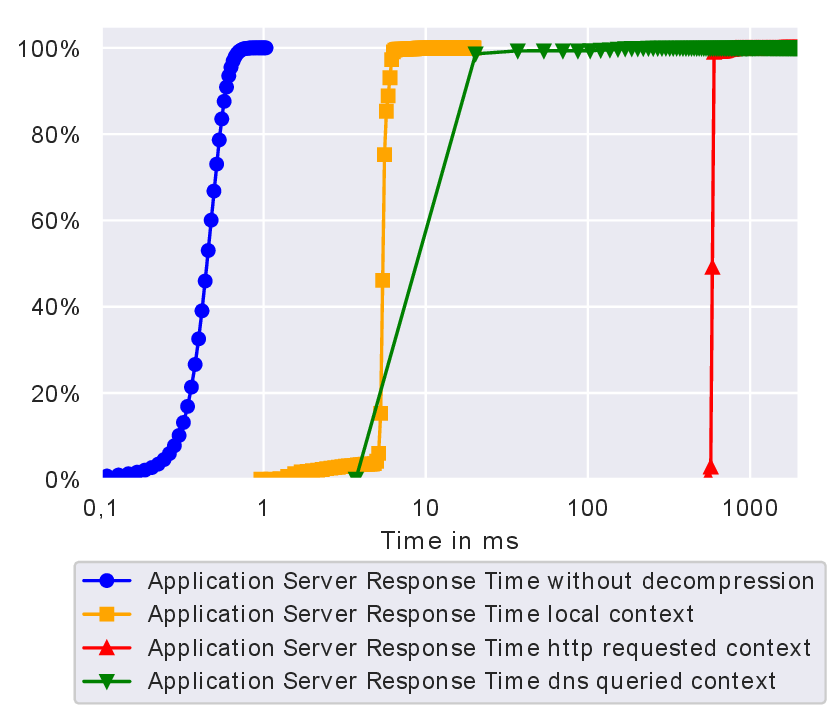}
    \caption{Cumulative distribution function of the Application Server Response Time $t1' - t0'$ (in \%) against (logarithmic scale) time in ms for all scenarios\label{srt_all}.}
\end{figure}

Fig. \ref{srt_all} illustrates the cumulative distribution functions of the Application Server-side Response Time $t1'-t0'$ with or without SCHC (cf. Fig. \ref{ex_diag_no_decomp} and \ref{ex_diag_local}) to show the order of magnitude of the sole decompression mechanism. We considered that a locally stored context file is used for the decompression. The curves show that integrating SCHC adds a few milliseconds to the operations necessary to work on the data independently to the possible delays added by the rule-querying mechanism.

%\begin{figure}[ht!]
%    \centering
%    \includegraphics[width=\linewidth]{SRT_w_wo_decomp.eps}
%    \caption{Cumulative distribution function of the Application Server Response Time $t1'-t0'$ (in \%) against time in ms for Scenarios 1 and 2 \label{srt_schc_int}}
%\end{figure}

Fig. \ref{srt_all} also shows the cumulative distribution functions of the Server-side Response Time $t1'-t0'$ for all the studied scenarios, thus including also context remote querying for the non-local solutions(HTTP, DNS).

We observe that the order of magnitude of the Server Response Time is for the worst case (HTTP-base mechanism) around 0.6s

Note that the DNS response time in our case (between 5 ms and 15 ms) is faster than the usual DNS response time due to DNS caching \cite{dns_caching} from our local network's DNS resolver. We keep interrogating our resolver with data it has already in its DNS cache so the DNS Response Time is cut down. In a wide LoRa deployment, this caching will remain, but considering the frequency with which the LoRa devices are expected to communicate on the network, the cache will probably be emptied from the necessary data. 

In order to provide a more realistic model to study the influence of adding DNS queries in an IoT system, we decided to gather additional data on DNS response time. We used RIPE Atlas \cite{atlas} which is a system that allows us to perform internet measurements through a set of probes available all over the world. While most of the probes are in Europe, we realized measurements asking for interrogations from all continents to test the responses for a single DNS query from multiple locations around the world. The measurements performed using RIPE Atlas allows us to determine the DNS Response Time in a more realistic case, and allows us to perform our query when it is certain that the DNS cache is expired.

\begin{figure}[ht!]
    \centering
    \includegraphics[width=\linewidth]{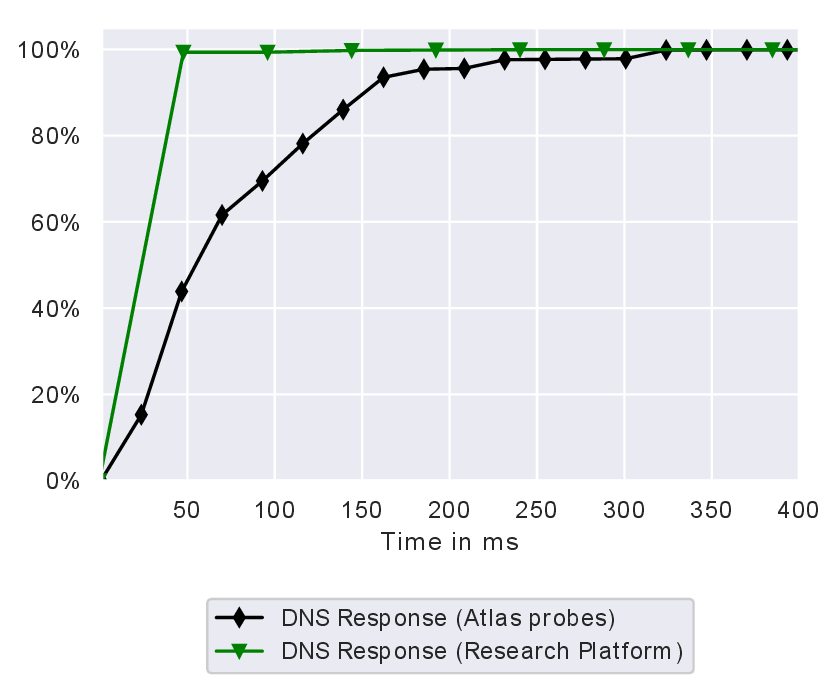}
    \caption{Cumulative distribution function of the DNS Response Time $t1'' - t0''$ (in \%) against time in ms for Scenario 3 compared and from RIPE Atlas \cite{atlas} Measurements \label{dns_resp}}
\end{figure}

Fig. \ref{dns_resp} provides a comparison between DNS Response Time for our DNS Queries and DNS Response Time obtained through measurements from RIPE Atlas interrogations. According to this figure, DNS Response will be slower in a real case than with our platform, but a time within 200ms is still perfectly viable with regards to the Response Times we measured for our platform.

\begin{figure}[ht!]
    \centering
    \includegraphics[width=\linewidth]{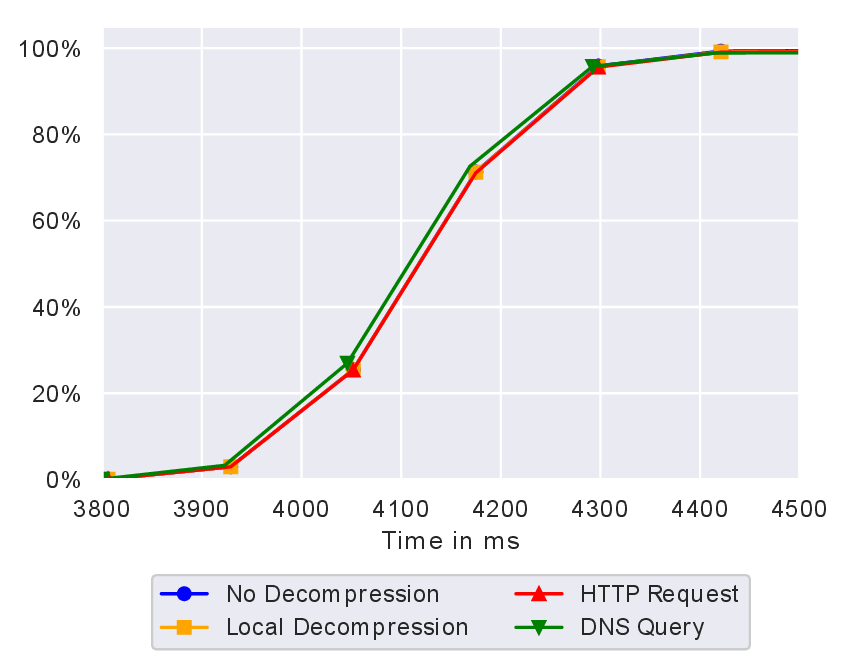}
    \caption{Cumulative distribution function of the Round Trip Time $t1 - t0$ (in \%) against time in ms for all scenarios (all the curves are superposed) \label{rtt}}
\end{figure}

\begin{figure}[ht!]
    \centering
    \includegraphics[width=\linewidth]{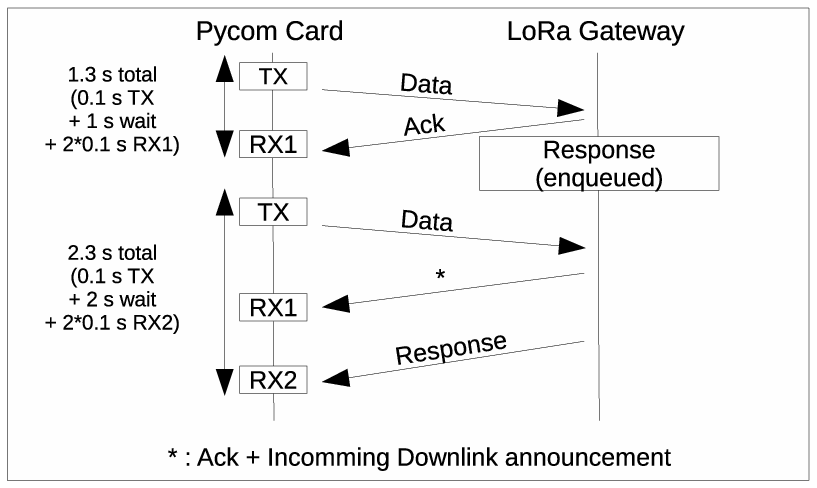}
    \caption{LoRa Communication Timing \label{ltw_rtt}}
\end{figure}

Fig. \ref{rtt} presents the measured Uplink Round Trip Time (From ED to Application Server, then back to ED) $t1 - t0$. We observe that it is at least about 3.9s for 99\% of the packets transmitted through our platform for all the studied scenarios. Considering the case of LoRa Class A devices \cite{lorawan}, a downlink frame from the gateway can only be sent during a given time interval called "receive window" (cf. \cite{lorawan} and \cite{draft_lpwan_rto}). The gateway implementation we are working with does not allow a frame to be transmitted to the device unless it has been en-queued before the gateway receives an uplink frame from the device (cf. Fig. \ref{ex_diag_all}). The last receive window is opened two seconds after the last uplink frame has been transmitted. It lasts twice the transmission time which depends on the SF. In our case, we use SF7 and our transmission time is around 100ms. For the majority of our measurements, our total measured  RTT is around 4.2s, Fig. \ref{ltw_rtt} illustrates how the gateway handles the data transmission to the device, which packet type are transmitted and which reception window is opened by the device to receive the data. Note that the implementation we use is ChirpStack, the OpenSource reference solution for LoRa platform. We would expect the same behaviour for class B devices whereas Class C would allow an immediate response and a shorter RTT.

%\begin{figure}[ht!]
%    \centering
%    \includegraphics[width=\linewidth]{Lora_Transmission_Window.eps}
%    \caption{LoRa Transmission/Reception Windows \label{ltw}}
%\end{figure}

Should we consider ourselves in a generic case with another gateway implementation, our goal would be to keep the Response Time as short as possible in order to allow the response to be transmitted to the device during one of these two receive windows. Our proposed mechanism introduces around a 650ms delay to Application Server Response Time leaving only 350ms of data handling. Using a DNS-only solution (i.e. the rule is already in cache as in Scenario 4) would also shorten this time to a 50 ms Application Server Response Time according to our platform. But Atlas measurements would show that this time is rather around 200 ms as illustrated by Fig. \ref{dns_resp}) which is still acceptable after introducing such a dynamic mechanism and considering our 1s window.

Problems may arise considering upper-layer protocols such as CoAP. This question is currently being asked at the IETF by C. Gomez and J. Crowcroft in their draft RTO considerations in LPWAN \cite{draft_lpwan_rto} for which the authors signal that "LoRaWAN policies may lead to U-RTT up to 282 seconds in the worst-case" (SF12). Working with SF7 measurements, we only observe a 4.1s mean RTT when considering the listening window used by the device to receive communication from the gateway. Actually, according to the \cite{rfc7252} CoAP message transmission has a default ACK\_TIMEOUT parameter which is set to 2 seconds. In this case, the ACK\_TIMEOUT has to be adjusted carefully to respect end-to-end delays.

\section{Conclusion}

Using SCHC to send IPv6 packets over LPWAN is proven to be an efficient way to take into consideration the scarcity of radio resources. We deployed all the components of a LoRaWAN infrastructure in order to build a SCHC-enabled LoRa network. Because of the expected large number of devices and the variety of possible things profiles, it seems necessary to envisage a mechanism to dynamically retrieve SCHC context rules. 

DNS is a global and well-known system which is a basic stepping stone when designing a dynamic system. Here we proposed a remote mechanism using DNS to retrieve rules signatures and allow a LoRaWAN infrastructure to react quickly to new devices communications or updated rules and eventually to retrieve the rules necessary to decompress data in a LoRa frame. This mechanism adds a small delay because of the RTT needed to query the DNS (and possibly a remote server as fallback mechanism to download rules). But this delay is acceptable considering the complete RTT of the LoRa specifications. Upper layers' timeout may have to be adjusted to take into account the corresponding added delay.

For our study, we focused on storing SCHC rule hashes using the DNS infrastructure. But hosting other device-specific information in the DNS such as device version, functionalities, status or expected battery lifetime is entirely possible. Provisioning these information in the DNS would allow for quick modification and easy propagation of the information to the customers with garanteed interoperability with any system and at a low cost.

\bibliographystyle{unsrt}
\bibliography{bibtex}

\section*{About}
\begin{spacing}{0.8}
{\footnotesize \textbf{$ ^{1}$Afnic} is a non-profit organization created in 1998 as a spin-off from the French Institute for Research in Computer Science and Automation (INRIA) which had hosted the pioneering work on the French Internet since the mid-1980s. Afnic manages the French TLD (.fr) and French Overseas TLDs, representing more than 3.5 million domain names. Its role forms part of a broader task of providing services of general interest, by contributing each day through the efforts of its teams and its members in order to provide an Internet that is secure and stable, open to innovation and in which the French internet community plays a key role.}

{\footnotesize \textbf{$ ^{2}$Telecom SudParis}, established in 1979 as the Institut national des télécommunications (INT), is a Grande École which belongs to the Institut Mines-Télécom (IMT) Group, Telecom SudParis has a recognized expertise in many fields such as : Applied Mathematics, Data Processing, Networks, Physics and Communications Technologies and Signal and Image Processing. Telecom SudParis conduct high-level research activities that address today’s major societal issues within the framework of the SAMOVAR CNRS Research Laboratory, in close collaboration with industry. Telecom SudParis is a founding member of "Institut Polytechnique de Paris", a group of higher education and research institutions that provide high-quality education and research services, and aiming to position itself as a French representative among the top institutions world-wide.}
\end{spacing}

\end{document}